\documentclass{classe}

\begin{document}

\title{Cognitive scale-free networks as 
a model for intermittency in
human natural language}

\author{Paolo Allegrini}

\address{ILC-CNR Area della Ricerca di Pisa, via
Moruzzi 1, 56010 Pisa, Italy\\E-mail: allegrip@ilc.cnr.it
}

\author{Paolo Grigolini}

\address{Dipartimento di Fisica, Universit\`a di Pisa and INFM, via
Buonarroti 2, 56127 Pisa Italy\\
Center for Nonlinear Science, UNT, P.O. Box 311427, Denton, Texas 76203-1427\\
IPCF-CNR, Area della Ricerca di Pisa, via Moruzzi 1,
56010 Pisa, Italy\\
E-mail: grigo@df.unipi.it
}

\author{Luigi Palatella}

\address{Dipartimento di Fisica, Universit\`a di Pisa and INFM, via
Buonarroti 2, 56127 Pisa Italy\\
E-mail: grigo@df.unipi.it   
}

\maketitle

\abstracts{
We model certain features of human language
complexity by means of advanced concepts borrowed from statistical
mechanics. Using a time series approach, the diffusion entropy method
(DE), we compute the complexity of an Italian corpus of newspapers and
magazines. We find that the anomalous scaling index is compatible with
a simple dynamical model, a random walk on a complex scale-free
network, which is linguistically related to Saussurre's {\em
paradigms}.  
The
model yields the famous Zipf's law in terms of the
generalized central limit theorem.}

\section{Introduction}

\begin{figure}[!h]
\begin{center}
\includegraphics[width=8.1 cm, height=5cm]{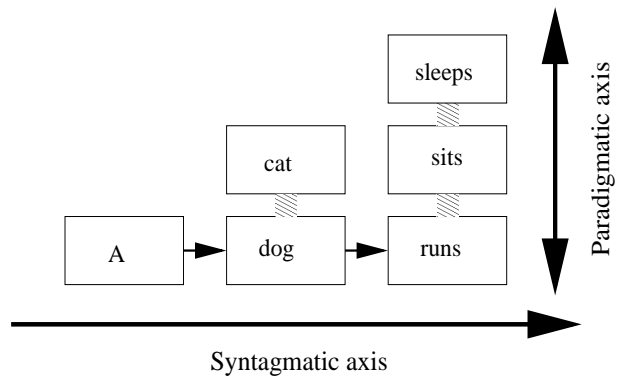}
\caption{\label{saussurre} Saussurre's dimensions. In this
example the first position in the syntagmatic axis is an article,
the second a noun and the third a verb in the third person.}
\end{center}
\end{figure}

Semiotics studies linguistic signs,
their meanings, and identifies the relations
between signs and meanings, and among signs.
The relations among signs (letters, words), are
divided into two large groups, namely the
syntagmatic and the paradigmatic, corresponding to 
what are called Saussurre's dimensions \cite{silverman}. 
These dimensions
are analogous to physical concepts like time and space.
One can grasp an understanding of them by looking at
Fig. \ref{saussurre}. The abscissa axis represents the syntagmatic
dimension, while the ordinate axis represents the paradigmatic
one. Along the abscissa grammatical rules pose constraints on how
words follow each other. This dimension is a temporal
one, with a casual order. An article (as ``a'' or ``the''), 
e.g., may be followed by
an adjective or a noun, but not by a verb of finite form. At a larger
``time-scale'', {\em pragmatic} constraints rule the succession of concepts,
to give {\em logic} to the {\em discourse}. 
The other axis, on the other hand,
refers to a ``mental'' space. The speaker has in mind a repertoire of
words, divided in many categories, which can be hierarchically
complex and refer to syntactical or semantic ``interchangeability''. 
Different {\em space}-scales of word paradigms can be associated to 
different levels of this
hierarchy. After an article, to follow the preceding example,
one can choose, at a syntactical level, among all nouns of a
dictionary. 
However, at a deeper level, semantic
constraints reduce the available words to be chosen. For instance,
after ``a dog'' one can choose any verb, but in practice 
only among verbs selected by semantic constraints
(a dog runs or sits, but does not read or smoke). The sentence 
``a dog graduates'', for instance, fits 
paradigmatic and syntagmatic rules behind Fig. 1,
but the semantics would in general forbid the production
of such a ``nonsensical'' sentence.

The two dimensions are therefore not quite orthogonal, and connect,
e.g., 
at a cognitive level.
The main focus of this paper is to show that this connection 
is in fact reproduced
at all scales. We shall also show that both dimensions are
{\em scale free} and that the complexity of linguistic structures in
both dimensions can be taken into account in a unified model, 
which is able to explain most statistical features of
human language, including, at the largest scales, 
the celebrated Zipf's law \cite{zipf}.

Zipf's law relates the rank $r$ of words to their frequency $f$
in a corpus. Remarkably, this does not mean that the probability of a word
is actually defined. In fact, a word may have a small or large
frequency depending on the genre of the {\em corpus} (i.e. a large
collection of written text) under study, and even two extremely large
corpora of the same type fail in reproducing the same word
frequencies. It is however remarkable that the occurrence of words is
such that {\em for any corpus and for any natural language} a property
emerge so that one finds only few frequent words and a large number of
words encountered once or twice. Let us define
word rank $r$, a property depending on the corpus adopted, as
follows. One assigns rank 1 to the most frequent word, rank 2 to the
second frequent one, and so on. Each word is uniquely associated to a
rank, and, although this number varies form corpus to corpus, one
always finds that

\begin{equation}\label{zlaw}
f \propto \frac{1}{r},
\end{equation}
This property means that {\em word frequencies
do not tend to well-defined probabilities}. We
assume that what can be defined is a ``probability of having a
frequency'' $P(f)$ for a randomly selected word. Operatively, one
measures $P(f)$ by counting how many words have a certain frequency $f$. In
Section III we show that a $P(f)$ compatible with the Zipf's law
can be derived from the model proposed herein, thus providing
our model an experimental support. For our scopes,
we assume a statistical mutual independence for the
occurrence of different {\em concepts}. This hypothesis is
appealing, since it means that every and each
occurrence of a concept makes entropy increase, thus
identifying the mathematical {\em information} (i.e. entropy) with the
common-sense information (i.e. the occurrence of concepts). Unfortunately,
a concept is not, {\em a priori}, a well defined quantity. 
Herein we assume
that concepts are represented by words (or better by lemmata) or by
groups of semantically similar words or lemmata\footnote{A lemma
is defined as a representative word of a class of words, having
different morphological features. For instance the word ``dogs''has
lemma ``dog'', and word ``sleeping'' has lemma ``sleep''.}.

Because of the mutual independence among different concepts, we can
extract from a single corpus as many ``experiments'' as the number of
concepts. For each experiment we select only one concept and we
mark the occurrence of the selected word or group of words
corresponding to this concept. 
For the analysis we use the recently
developed Diffusion Entropy (DE) method, which is able to identify
whether a marker is a ``real event'', i.e. it carries maximal information, and
to extract the scaling properties of the language dynamics. We show
here that anomalous scaling (different from Brownian motion) is an
indication of long-range correlations of the series, and that in fact
these properties are well measured by the DE, even if the marker is
not identified with absolute precision.

The overall dynamics, given by the flow of concepts over
time, experimentally mirrors the dynamics of intermittent dynamical systems,
like the Manneville's Map: These systems have
long periods of quiescence followed by bursts of activity. This
variability of waiting times between markers of activity is responsible
for long-range correlations \cite{manneville}.

The second aspect of the paper is the connection between 
space and time complexity, and its application to
linguistics. We will assume that atomic concepts
exist and represent nodes of a complex network, connected by
arcs representing, when existing, semantic {\em associations} 
between a concept and another. We assume that our markers are 
actually defined as a group
of neighboring nodes. We then assume that language can be
produced by a random walker, ``associatively'' traveling
from concept to concept. The scale-free properties of the
network, independently measured by our research group,
provides a bridge to understand the intermittent dynamics
earlier described. In this unified model the network
is a representation of Saussurre's paradigms, whose complexity 
mirrors the syntagmatic one in the asymptotic limit.  

\section{DE and concepts} 

Let us review the DE method
\cite{giacomo,giacomo2,giulia1,terremoti}. In synthesis, one defines a
``marker'' on a time sequences, and studies the probability $p(x;t)$
of having a number $x$ of markers in a window of length $t$. This
statistical analysis is done by moving a window of length $t$ along
the sequences, counting how many times one finds $x$ markers inside
this window, and dividing this number by the
total number  $N-t+1$ of windows of size $t$, where
$N$ is the total length.

Having large number values for $x$ and $t$, we can
adopt a continuous approximation. Moreover,
in the ergodic and stationary condition, a scaling relation is
expected, namely

\begin{equation}
\label{scaling}
p(x;t)=\frac{1}{t^{\delta}} F \left( \frac{x-wt}{t^{\delta}} \right),
\end{equation}
where $w$ is the overall marker density, $\delta$
is the {\em scaling index} and $F$ is a function. 
If $F$ is the Gauss function, $\delta$ is the
known Hurst index, and if the further condition $\delta=0.5$ is
obeyed, then the process is said to be Poissonian, and the dynamics of
$x$ is called ``Brownian motion''. If this condition applies, there is
no long-range memory regulating the occurrence of markers in time.

It is straightforward to show that
$S(t)=\int_{-\infty}^{\infty}dx p(x;t) \ln p(x;t)$,
namely the Shannon Information,
with condition (\ref{scaling}), leads to

\begin{equation}
S(t)=k + \delta \ln t,
\end{equation}
where $k$ is a constant. The evaluation of the slope
according to which $S$ increases with $\ln t$ provides
therefore a measure for the anomalous scaling $\delta$.

Let us briefly mention what we know about applying DE to time
series with known long-range correlation. We construct an artificial 
series by letting $\xi_i=1$
(this means that we find the marker at the $i$th position), or $\xi_i=0$
(the $i$-th sign is not a marker). We then assume ``informativity''
for the marker (markers are then called ``events''), namely that the
distance between a ``1'' and the successive does not depend on the
such previous distances. Then, if the distances $t$ between events are
distributed as
\begin{equation}
\label{model}
\psi(t) = (\mu - 1) \frac{T^{\mu-1}}{(t+T)^{\mu}}
\end{equation}
($\psi(t) \sim t^{-\mu}$ asymptotically is a sufficient condition),
then the theory based on continuous-time random walk and on the
generalized central-limit theorem yields for $p(x;t)$ a truncated L\'evy
probability distribution function (PDF) \cite{giacomo}. 
DE detects the
scaling $\delta$ of the central part, namely

\begin{equation}
\label{deltamu}
\delta=\frac{1}{\mu-1}  \:\: \mbox{if} \:
2<\mu<3,\:  \delta=0.5 \:\:  \mbox{if} \: \mu>3.
\end{equation}
The condition $2<\mu<3$ means long-range correlation, since for
truncated L\'evy PDFs asymptotically 
$\langle x^2 (t) \rangle - \langle x (t) \rangle^2 \propto t^{4 -\mu}$
and therefore
the
correlation function decays as $t^{\mu-2}$.
Note that the decay of this correlation function is
non-integrable, yielding an infinite correlation time.
The theory rests on a dichotomous $\xi$, and experimentally
this means the presence or absence of a certain
marker. One may, for instance look for a certain letter, so that the
time is the ordinal number of the typographical characters in the
text. 
As later shown,
we have better results by looking at lemmata, where the ``time'' is
the ordinal number of words. We shall show that, with a good choice 
of {\em semantic} markers, Eq. (\ref{model}) is a good model for 
concepts dynamics in natural language.

Eq. (\ref{deltamu}) \cite {giacomo} rests on uncorrelated waiting times
between events. This means that if two markers are separated by
intervals of words of duration $\tau_k$ (the distance in words between
the $k$-th and the $k+1$-th occurrence of the marker) then
$\langle \tau_i \tau_j \rangle \propto \delta_{ij}$, where $\delta_{ij}$
is the Kroeneker delta. Under these conditions each event
carries the same amount of information. The statistical independence
between the $\tau_k$ intervals means that the information carried by
the events is maximal for a given waiting time distribution
$\psi(\tau)$. In a linguistic jargon, we can say that if in a corpus
we find a marker (e.g. a list of
words) such that $\delta \approx 1/(\mu - 1)$ then this marker is {\em
informative} in that corpus. For didactical purposes, we shall see
that certain markers, e.g. punctuation marks, are not {\em real
events}, but are rather modeled by a Copying Mistake Map (CMM)
\cite{maria}.
This means that discourse complex dynamic is such that the
punctuation marks actually carry long-range correlations,
and anomalous scaling in the PDF, while the waiting times between
such marks are correlated. Punctuation marks are not
informative. Their complexity is just a projection of a complexity
carried by ``concept dynamics''.

\subsection{The CMM and non-informative markers}
The Copying Mistake Map (CMM) \cite{maria} is a model originally introduced to
study the anomalous statistics of nucleotides dispersion in coding
and non-coding DNA regions.  The CMM is a combination of two
sequences: We have an ``original'' time sequence like e.g. the
long-range-correlated series earlier discussed, corresponding to
the waiting time distribution (\ref{model}). Then, for any $\xi_i$
we either leave it unchanged with probability $\epsilon$ or change it
with a completely random value with probability $1-\epsilon$ (copying
mistake).

The resulting waiting time distribution decays exponentially, since
the probability of finding a 1 after a time t from the preceeding
one, is given by two terms. This is because the 1 can be
associated to two kinds of origin: it may be an ``original'' 1, 
or an original ``zero'' flipped by the copying mistake. 
We can write the ``experimental'' waiting-time distribution
$psi_{exp}(t)$, in terms of $psi_{corr}(t)$ of the mentioned
long-range-correlated model (\ref{model}), and of $psi_{rand}(t)$ of the
Poissonian copying process, namely

\begin{equation}
\psi_{exp}(t)=\psi_{rand}(t) \Psi_{corr}(t)+ \Psi_{rand}(t) \psi_{corr}(t),
\end{equation}
where $\Psi(t)\equiv \int_{t}^{\infty}dt' \psi(t')$ and
$\Psi_{rand}(t)\equiv \int_{t}^{\infty}dt' \psi_{rand}(t')$,
and

\begin{equation}
\psi_{rand}(t)=\ln \left( \frac{2}{1 - \epsilon} \right) \cdot \left( \frac{2}{1-\epsilon} \right)^{-t}.
\end{equation}
Since $\psi_{rand}(t)$ and consequently $\Psi_{rand}(t)$ decay as an
exponential function, so it does, in the asymptotic limit,
$\psi_{exp}(t)$.
What about the DE curve?
The theory predicts \cite{scafettalatora} a random ($\delta=0.5$)
behavior for short times, a knee, and a slow transition to the totally
correlated behavior. An example of CMM is given by punctuation marks
in Natural Language.  We choose punctuation marks as markers for an
Italian corpus of newspaper and magazines, of more than 300,000 words
length, called {\em Italian Treebank} (hereafter TB). In
this experiment we look at words, and we put a 0 for every word which
is not a punctuation mark, and a 1 when we find such a mark (full
stops, commas, etc.). A sentence like ``Felix, the cat, sleeps!'' is
therefore transformed into ``0 1 0 0 1 0 1''.

Fig. 2 shows that this markers lead to a time series with all the
earlier exposed features of a CMM. This means that the waiting times
$\tau_k$ are correlated, and therefore punctuation marks are not
events. Notice however that an asymptotic anomalous $\delta$ is
detected by DE, and therefore there is a long-range correlation in the
text, which may be carried by some other more informative marker.


\begin{figure}
\begin{center}
\large
\vspace{35pt}{\bf\hspace{-6.3cm}(a)\vspace{-35pt}} \\
\includegraphics[width=10 cm,height=7.5cm ]{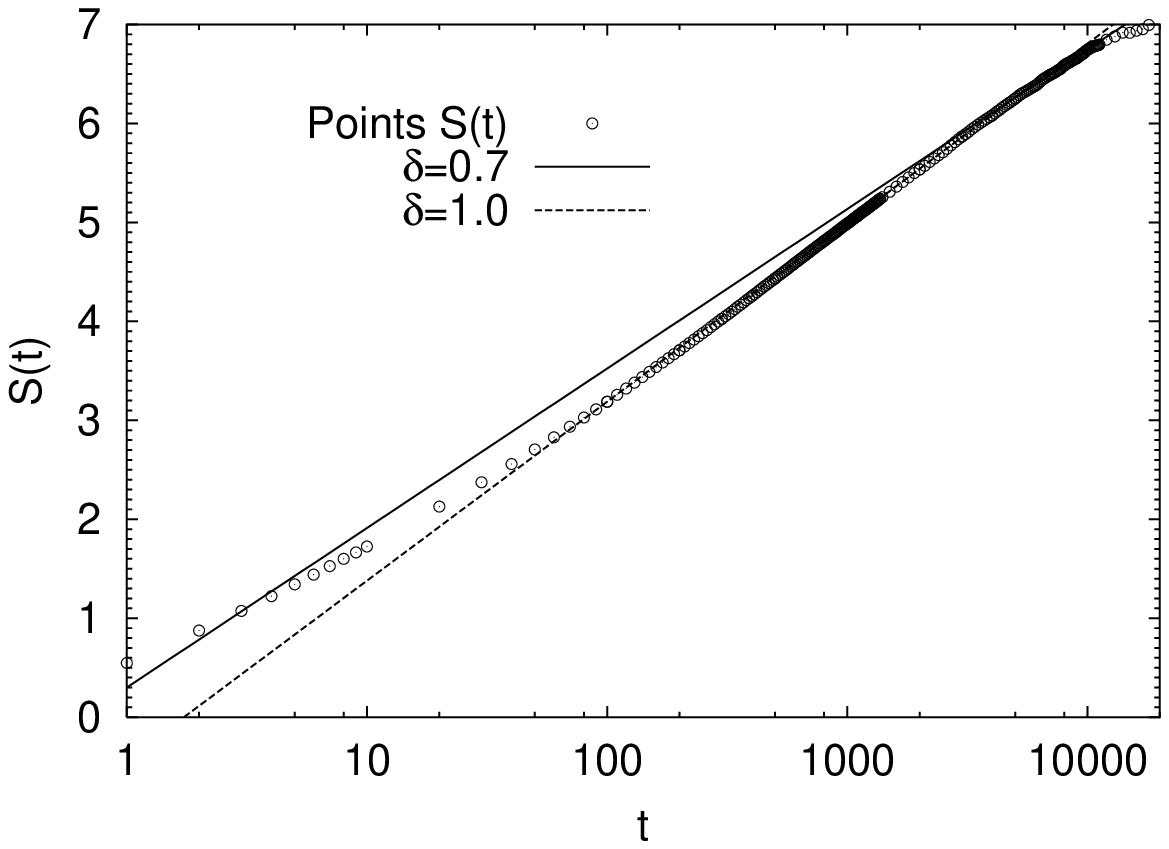}\\
\vspace{25pt}
{\bf\hspace{-2.6cm}(b)\hspace{5.5cm}(c)\vspace{-25pt}}
\normalsize \\
\includegraphics[width=6.0 cm,height=5.2cm]{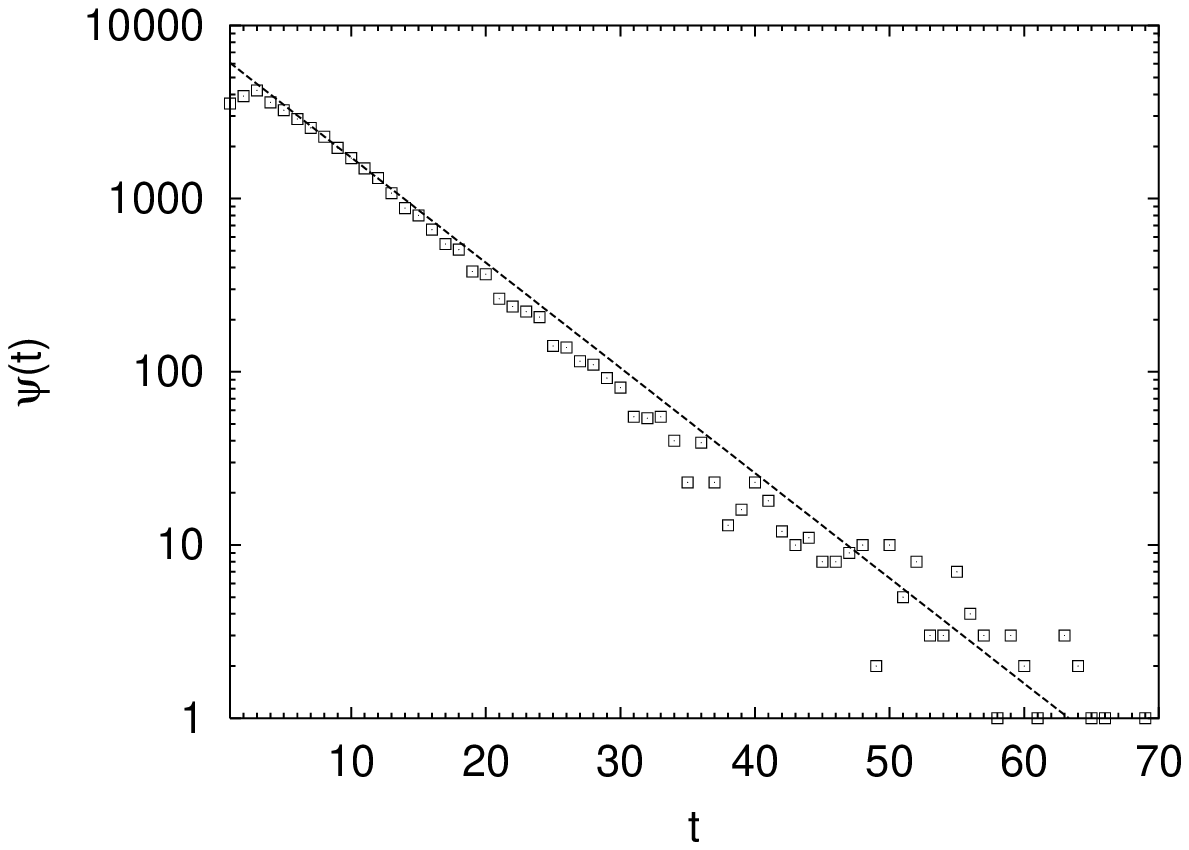}
\includegraphics[width=6.5 cm,height=5.2cm]{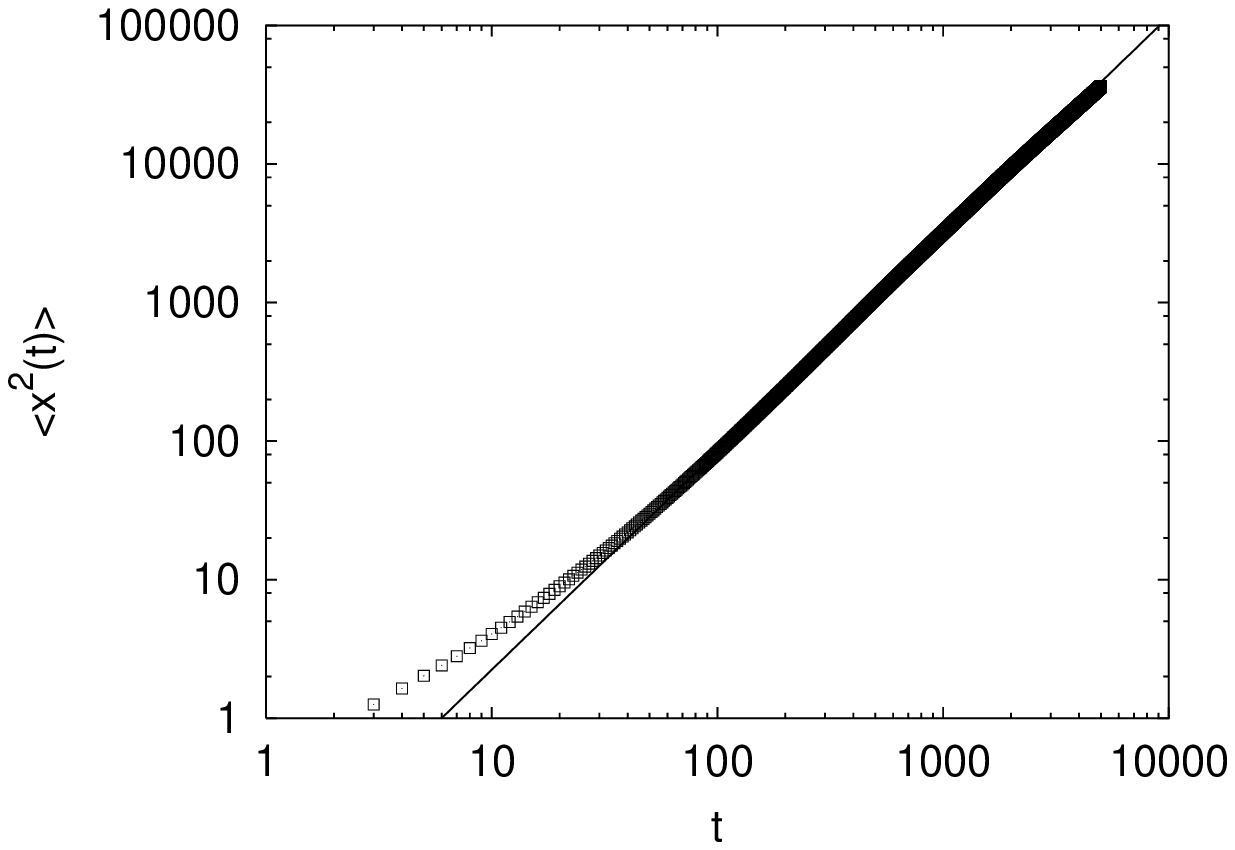}
\caption{ (a) Diffusion entropy for punctuation marks. The fit for the
asymptotic limit (solid line) yields a $\delta =0.7$. The dashed line
marks a transient regime with $\delta=1$. (b) Non-normalized
distribution of waiting times for punctuation marks, namely counts of
waiting times of length t between marks in TB. The expression for the
dashed line fit is $7000 \cdot \exp (-t/7.15)$. (c) Second moment
analysis for punctuation marks. The expression for the solid line fit
is $0.06\cdot t^{1.57}$. Notice that $1.57 \approx 3-1/0.7$, namely
the expression $H=3-1/\delta$ of Ref. 9 for L\'evy
processes stemming from CMM's is verified.}
\end{center}
\end{figure}



\subsection{Concepts as informative markers}

More experiments, not reported here, show that the CMM
behavior is typical for many characters, and are shared
by all the letters of the alphabet, with a $\delta \approx 0.6$.
Passing from a ``phonetic'' (in Italian we can
assume that alphabetic characters mirror the phonetic)
to a morpho-syntactic level is linguistically interesting.
To do so, a text has to be lemmatized and tagged with respect
to its part of speech. After this procedure we can identify
as a marker the occurrence of a certain part of speech
(e.g. article, adverb, adjective, verb, noun, preposition, numeral,
punctuation etc.). For instance, the sentence ``Felix, the cat, sleeps!''
is now transformed into ``N P R N P V P'', where N, P, R and V
stand for nouns, punctuation, article and verb. If we select
the occurrence of verb as a marker, then we have 
``0 0 0 0 0 1 0''. Fig 3 shows the result of this experiment
for verbs and for numerals. We notice that we have a similar
behavior for the DE, and a completely different behavior for
the evaluation of the waiting-times distribution $\psi(t)$
(where $t$ is the number of words between markers). We notice that
DE reveals a long-time correlation, while, $\psi(t)$ shows an exponential
truncation at long times. However, in the case of numerals
we find a large transient with a slope $\mu_{numerals} < 2$,
and therefore a non-stationary behavior. This is in fact due
to the uneven distribution of numerals in the corpus, since they
are encountered more often in the economic part of the Italian
newspapers. However, this still unsatisfactory result for numerals
reveals that this kind of markers is more informative than a phonetic
one or than the presence of verbs. 
This is linguistically interesting, since numerals denote
a part of speech, but also a ``semantic class''.

\begin{figure}[hbt]
\begin{center}\vspace{30pt}
\large
{\bf\hspace{3cm}(a)\hspace{6.5cm}(b)\vspace{-30pt}}
\normalsize \\
\includegraphics[width =5.5 cm,height=5cm]{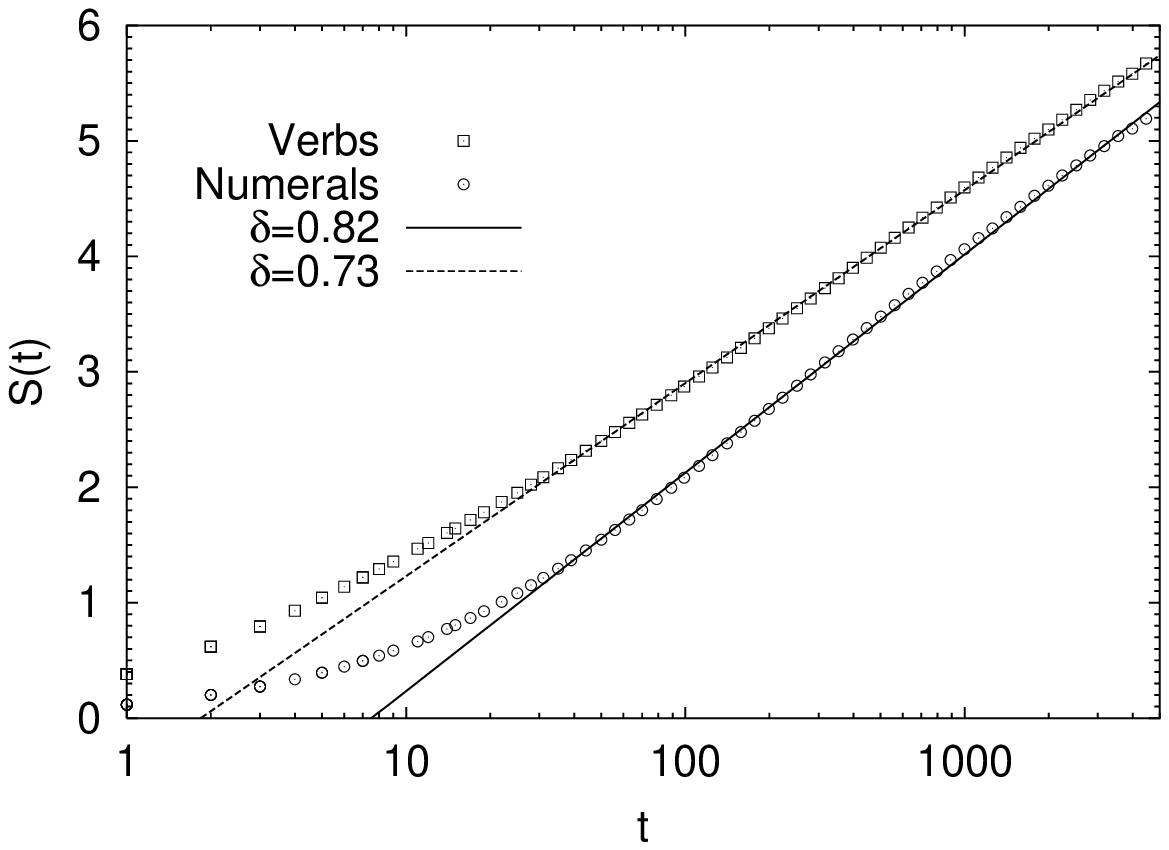}
\includegraphics[width= 6.5 cm,height=5cm]{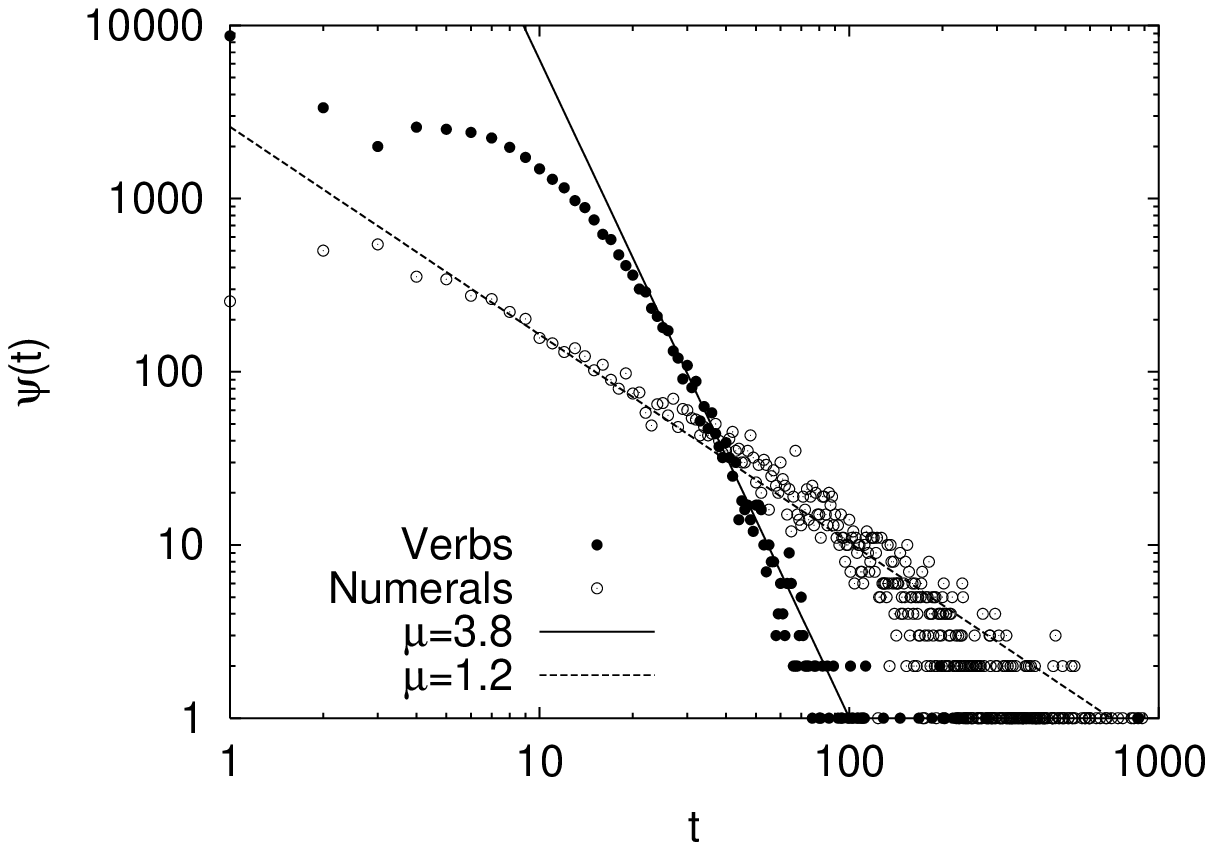}
\caption{a) DE for verbs (squares) and numerals (circles). The dashed
line is a fit for the verbs, with $\delta=0.73$, while the solid line
is a fit for the numerals, with $\delta=0.82$. b) $\psi(t)$ for verbs
(black circles) and numerals (white circles). The dashed line is a fit for the
numerals, with $\mu=1.2$, while the solid line is a fit for the
verbs, with $\mu=3.8$.}
\end{center}
\end{figure}


We are therefore led to suppose that informative markers are the ones
associated with a semantically coherent class of words. This is
however a problem, since every single concept is too rare in a balanced
corpus (a long text with a variety of genres). The next level of our
exploratory search for events is therefore to look at the occurrence of
``salient words'' in a specialistic text. Such a corpus has been made
available as the Italian corpus relative to the European project
POESIA \cite{poesia}. POESIA is a European Union funded project whose
aim is to protect children from offensive web contents, like,
e.g. pornography in WWW URLs. Salient ``pornographic'' words were
automatically extracted by comparing their frequency in an offensive
corpus, with respect to the balanced TB-corpus. The
definition adopted was
\begin{equation}
s(l)=\frac{f_{EC}(l)-f_{TB}(l)}{f_{EC}(l)+f_{TB}(l)},
\end{equation}
where $f_{EC}(l)$ is the frequency, in the erotic corpus, of the lemma
$l$, and $f_{TB}(l)$ is the same property in the reference Italian
corpus (Italian Treebank). Salient lemmata were automatically chosen
as the 5\% with the highest value of $s$. 
Notice that in this experiment all ``dirty'' words are not
taken into consideration, because they do not appear in the
reference corpus, and therefore $s$ cannot be properly defined. 
However an offensive metaphoric use of
terms is in fact detected, leading to a completely
new way to automatic text categorization and filter \cite{poesia},
using a method, based on DE analysis, called CASSANDRA \cite{granada}.

Salient words were therefore used as markers for our analysis, as
earlier described. The results are shown in Fig. 4, clearly
showing that {\em in a specialized corpus, salient words of this
genre, pass the test of informativeness}. Salient words, and plausibly
words in general, are therefore distributed like markers generated by
an intermittent dynamical model, with $\mu \approx 2.1$ and, in
agreement with (\ref{deltamu}), $\delta \approx 1/(\mu-1) = 0.91$. We
see in the next section how this behavior is plausibly connected with a
topological complexity at the paradigmatic level, and in Section III we
derive the Zipf's law from the resulting model.

\begin{figure}[htb]
\begin{center}\vspace{30pt}
\large
{\bf\hspace{3cm}(a)\hspace{6,5cm}(b)\vspace{-30pt}}
\normalsize \\
\includegraphics[width = 5.5 cm, height=5 cm]{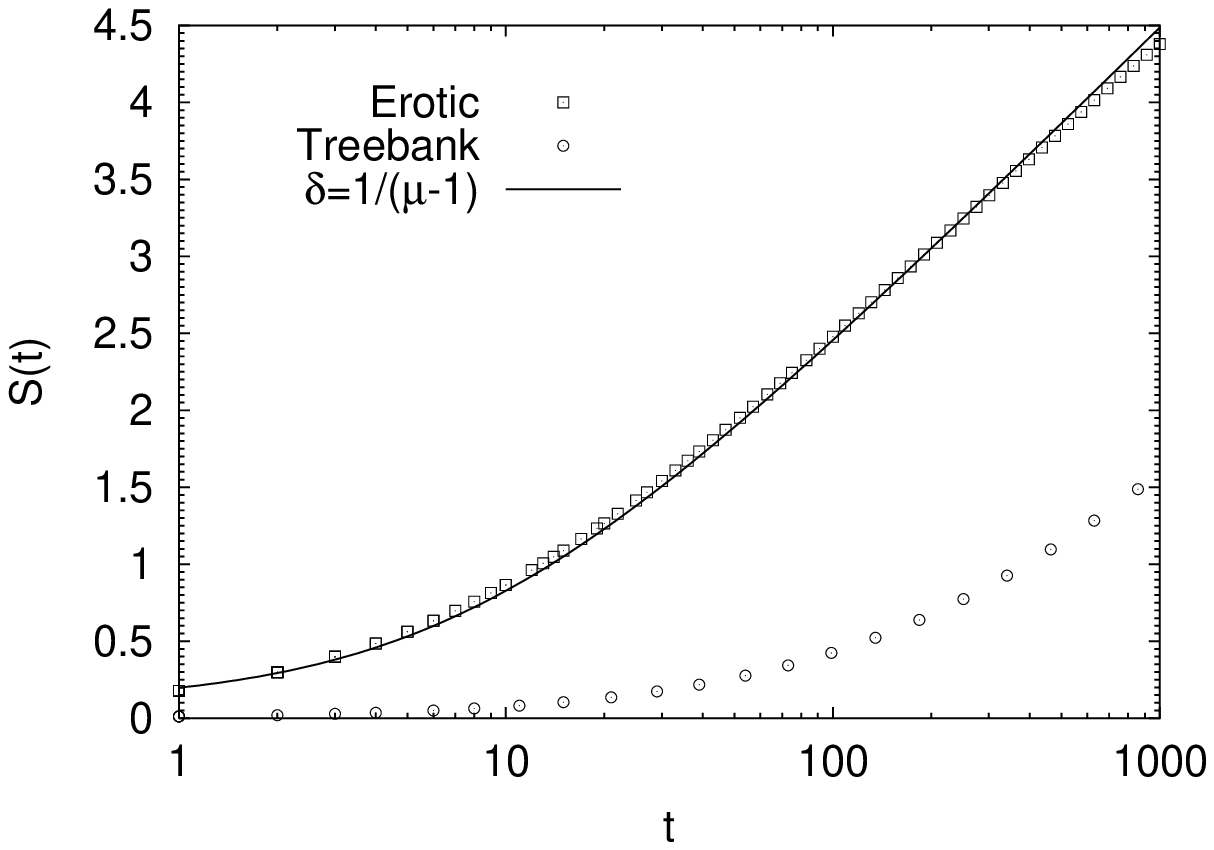}
\includegraphics[width = 6.5 cm,height=5 cm ]{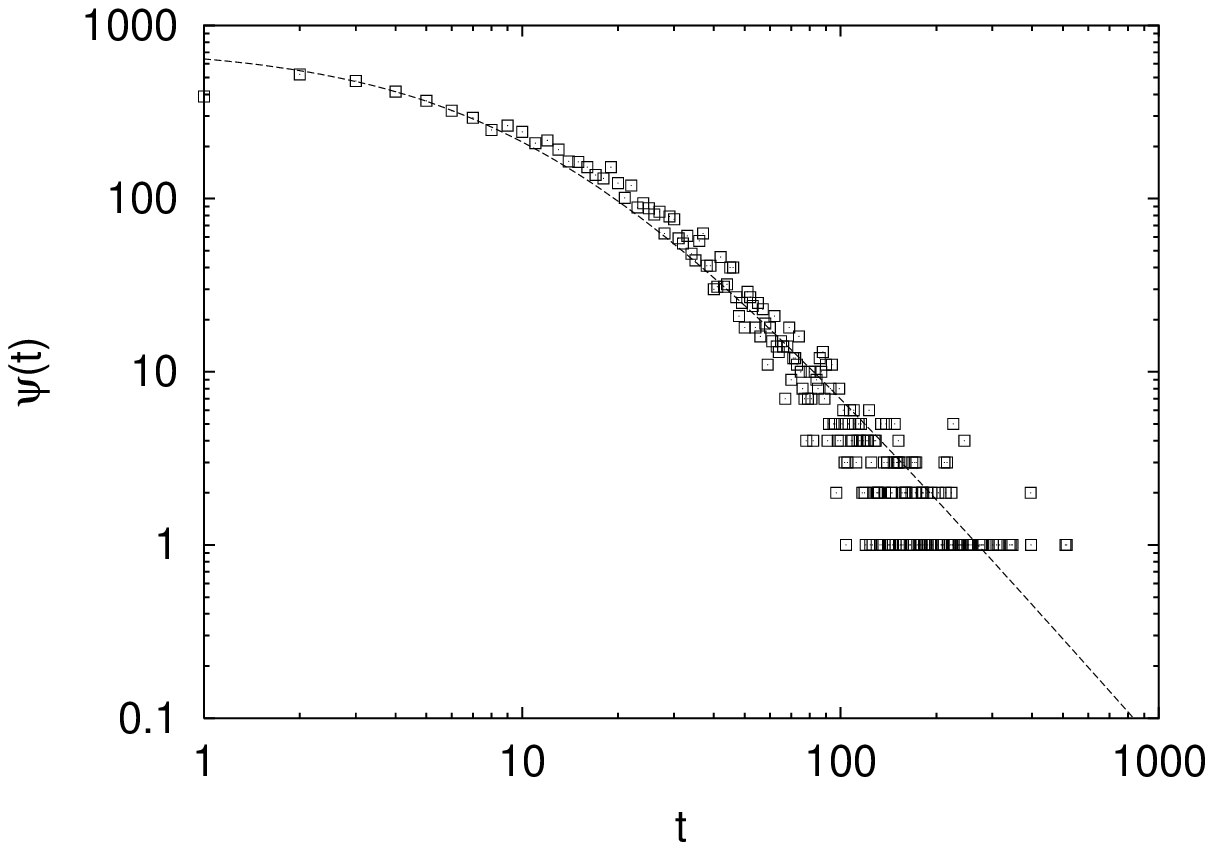}
\caption{a) DE for salient ``erotic'' words for a corpus of erotic
  stories and offensive web pages (squares), and for the Italian
  reference corpus (circles). The solid line is a fit with expression
  $S(t)=k+\delta\ln(t+t_0)$, where the additional parameter $t_0$ is
  added to the original Eq. (3) to take transients into account and to
  improve the quality of the fit, yielding $\delta=0.91$ b)
  Non-normalized waiting time distribution for salient ``erotic''
  words for a corpus of erotic stories. The expression for dashed line
  fit is $14000\cdot(12.0 + t)^{-2.1}$, yielding $\mu=2.1$.}
\end{center}
\end{figure}



\section{Scale-free networks, intermittency and the Zipf's law}

In this section we build a cognitive model for
connecting structure and dynamics. Allegrini et al. 
\cite{ilc} identified semantic classes in the Italian corpus, by
looking at paradigmatic properties of interchangeability
of classes of verbs with respect to classes of nouns. They defined
``superclasses'' of verbs and nouns as ``substitutability
islands'', namely groups of nouns and verbs sharing the properties
that in the corpus you find each verb of the class co-occurring,
in a context, with each noun of the class \cite{ilc}. This is precisely
a direct application of the notion of ``paradigm''. Let
us call $p_v(c)$ and $p_n(c)$, respectively, the number
of verbs or nouns belonging to a number $c$ of classes.
They found that
\begin{eqnarray}
p_v(c) \propto \frac{1}{c^{1+\eta}}\nonumber \\
p_n(c) \propto \frac{1}{c^{1+\eta}},
\end{eqnarray}
where $\eta$ is a number whose absolute value is (much) smaller than $1$.

On the same line, other authors \cite {smallworldenglish} found a
``small world'' topology \cite{smallworld}, by looking at the number
of synonyms in an English thesaurus, for each English lemma. We can
therefore assume that this kind of structure is general for any
language.  Let us therefore imagine that the paradigmatic structure of
concepts is a scale-free network and consider a random walk in this
``cognitive space''.  Let us make the following assumptions:\\
{\em
1) The statistical weight of the $i$-th node is $\omega_i \sim c_i$;\\
2) Ergodicity, and therefore that the characteristic recurrence
  time is 
$\tau_i \sim c_i^{-1}$;\\
3) The same form for all nodes,$\psi_i(t)=(1/\tau_i) 
F \left( -t/\tau_i \right) $ (e.g. $F(x)=\exp(-x)$).\\
}
Now we imagine that selecting a {\it concept} means selecting a few
neighboring nodes. This collection of nodes, due to the scale-free
hypothesis,
shares the same scaling properties
of the complete scale free network, namely $p(c) \sim c^{-\nu}$.
Therefore we have that
\begin{eqnarray}
\psi_{concept}(t)=\sum_{i} \omega_i \psi_i (t) \propto \sum_i c_i^2
e^{-c_i t} 
\approx \int dc  c^2 e^{-ct} \frac{1}{c^{\nu}} \sim
\frac{1}{t^{3-\nu}}.
\end{eqnarray}
We recovered the intermittent model (\ref{model}).

Let us now make the exercise of deriving the Zipf's law
$f \propto r^{-a}$,  with $a$ close to unity.
Let us define a probability of frequency $P(f)$

\begin{equation}\label{iacobian}
P(f)df = prob(r) dr \Longrightarrow P(f) \sim f^{-\frac{a+1}{a}}
\end{equation}
Next, let us notice that $P(f)$ must be a stable distribution.
In fact, the Zipf's law is valid for every corpus. In particular
if it is valid for corpus $A$ and for corpus $B$, it is valid
also for the corpus $A+B$ where $+$ means the concatenation
of corpora. If we continue with concatenating we will have
a corpus

\begin{center}
Total Corpus = Corpus A + Corpus B + ... 
\end{center}
and we write the frequency of a word in the total corpus, $f_{tot}$ is written
in terms of the single frequencies $f_1$, $f_2$, $\dots$, and total
lengths $N_1$, $N_2$, $\dots$ of the single corpora
\begin{eqnarray}
f_{tot} = \frac{f_1+f_2+ \cdots }{N_1+N_2+\cdots} 
= \frac{1}{\sum_i N_i}\sum_i f_i, 
\end{eqnarray}
i.e., the Generalized Central Limit Theorem \cite{GCLT} applies.  This
means that the probability of frequency $P(f)$ is a L\'evy
$\alpha$-stable distribution. This probability of finding $f$
occurrences of a word in a corpus of a given length can be identified
with $p(x;t)$ of Section II, if we take into consideration the
parameter $t$. We have earlier noticed that $p(x;t)$ in language is
L\'evy process, with $\delta \sim 1$, and therefore with a tail $P(f)
\sim f^{-2}$. In other words through (\ref{iacobian}) we recover
(\ref{zlaw}) i.e. the Zipf's law.

\section{Conclusions}

In this paper we have shown that a cognitive process governing human
language may be identified, and that it has a complexity both in the
syntagmatic and in the paradigmatic axis. The scaling properties of
both axes are related to each other, and are reflected by the celebrated
Zipf's law.  This study was conducted using Italian written corpora,
but decades of studies on the generality of the Zipf's law lead us to
suppose that our results are language independent, and that the
language complexity that we are revealing is genuine and important. In
fact, for any concept, we have a scaling index associated with an
intermittent dynamical model that rests at the border between
ergodicity and non-ergodicity, since the Zipf's law is theoretically
consistent with $\delta = 1$.  Moreover, in a
specialistic test we see a tendency to drift, for salient words,
towards ergodicity ($\delta \approx 0.91$ in the reported
experiment). This behavior can be interpreted as the balance
between two opposite needs for human language, namely {\em
  learnability}, i.e. the possibility for a child to learn a language by
examples, and {\em variability}, to explore an infinite cognitive
space. 

We propose as a future work to study language complexity in children
during learning years, and in psychopathological subjects. We
imagine, if the theory presented herein is validated by
more extensive work, that the simple study of the individual
Zipf's laws can provide a reasonable non-invasive diagnostic method 
for certain mental diseases.

From a Language Engineering point of view, this study provides a
theoretical background for a completely new strategy of automatic text
categorization. A prototype is being implemented as a semantic filter
\cite{poesia}. We think that the proposed test for informativeness for a set of
markers can also be important for many exploratory studies in time series
analysis. For instance, it may become important to identify
crucial semantic markers in a flow of data.


\end{document}